\documentclass[preprint,nofootinbib,showkeys,showpacs,tightenlines,fleqn]{revtex4}   
\usepackage{graphicx}  
\begin{document}      
\preprint{PNU-NTG-01/2006} 
\preprint{PNU-NURI-01/2006} 
\title{QCD condensates with flavor SU(3) symmetry 
breaking from the instanton vacuum}   
%================================================= 
\author{Seung-il Nam} 
\email{sinam@pusan.ac.kr} 
\affiliation{Department of 
Physics and Nuclear Physics \& Radiation Technology Institute (NuRI),    
Pusan National University, Busan 609-735, Republic of Korea} 
%===================================== 
\author{Hyun-Chul Kim} 
\email{hchkim@pusan.ac.kr} 
\affiliation{Department of 
Physics and Nuclear Physics \& Radiation Technology Institute (NuRI), 
Pusan National University, Busan 609-735, Republic of Korea} 
\date{December 2006} 
%===================================== 
\begin{abstract} 
We investigate the effects of flavor SU(3)-symmetry breaking on the  
quark, gluon, and mixed quark-gluon condensates, based on the nonlocal  
effective chiral action from the instanton vacuum.  We take into 
account the effects of the flavor SU(3) symmetry breaking in the 
effective chiral action, so that the dynamical quark mass depends on 
the current quark mass ($m_f$).  We compare the results of the present 
approach with those without the current quark mass dependence of the  
dynamical quark mass.  It is found that the result of the quark 
condensate is decreased by about 30 \% as $m_f$ increases to 200 MeV, 
while that of the quark-gluon mixed condensates is diminished by about 
$15\%$.  We obtain the ratios of the quark and quark-gluon mixed 
condensates, respectively: $[\langle \bar{s}s\rangle/\langle  
\bar{u}u\rangle]^{1/3}=0.75$ and 
$[\langle\bar{s}\sigma_{\mu\nu}G^{\mu\nu}s 
\rangle/\langle\bar{u}\sigma_{\mu\nu}G^{\mu\nu}u\rangle]^{1/5}=0.87$. 
It turns out that the dimensional parameter   
$m^2_0=\langle\bar{q} \sigma_{\mu\nu}G^{\mu\nu}q\rangle/ 
\langle\bar{q}q\rangle =1.60\sim 1.92\,{\rm GeV}^2$.   
\end{abstract} 
\maketitle 
%====================>>>>> 
%\section{introduction} 
%====================<<<<< 
{\bf 1.} Understanding the QCD vacuum is very complicated, since both 
perturbative and non-perturbative fluctuations come into  
play.  In particular, the quark and gluon condensates, being the  
lowest dimensional ones, characterize the non-perturbative structure of 
the QCD vacuum.  The quark condensate is identified as the order 
parameter for spontaneous chiral-symmetry breaking (S$\chi$SB) which 
plays an essential role in describing low-energy phenomena of hadrons: 
In the QCD sum rule, these condensates come from the operator product  
expansion and are related to hadronic 
observables~\cite{Shifman:1978bx}, while in chiral perturbation theory    
($\chi$PT), the free parameter $B_0$ is introduced in the mass term of 
the effective chiral Lagrangian at the leading 
order~\cite{Gasser:1983yg} measuring the strength of the  
quark condensate~\cite{Gell-Mann:1968rz}.  On the other hand, the 
gluon condensate is not the order parameter but measures the vacuum 
energy density~\cite{Shifman:1978bx}, which was first estimated by the  
charmonium sum rule~\cite{Shifman:1978by}.  
 
While the quark and gluon condensates are well understood 
phenomenologically, higher dimensional condensates suffer from 
large uncertainty.  Though it is still possible to estimate 
dimension-six four-quark condensates in terms of the quark condensate 
by using the factorization scheme which is justified in the large 
$N_c$ limit, the dimension-five mixed quark-gluon condensate is not 
easily determined phenomenologically.  In particular, the mixed 
condensate is an essential parameter to calculate baryon 
masses~\cite{Ioffe:1981kw}, exotic hybrid 
mesons~\cite{Balitsky:1986hf}, higher-twist meson distribution 
amplitudes~\cite{Chernyak:1983ej} within the QCD sum rules. 
Moreover, the mixed quark condensate plays a role of an   
additional order parameter for S$\chi$SB, since the quark chirality 
flips via the quark-gluon operator.  Thus, it is naturally expressed 
in terms of the quark condensate: 
\begin{equation} 
\langle\bar{\psi}\sigma^{\mu\nu}G_{\mu\nu } 
\psi\rangle = m_0^2 \langle\bar{\psi}\psi\rangle    
\end{equation} 
with the dimensional parameter $m_0^2$ which was estimated in various  
works~\cite{Ioffe:1981kw,Belyaev:1982sa,Dosch:1988vv,Dorokhov, 
Polyakov:1996kh,Doi:2002wk} and with the definition 
$G_{\mu\nu}=G^a_{\mu\nu}\lambda^a/2$.  
 
The instanton picture allows us to study the QCD vacuum 
microscopically.  Since the instanton picture provides a natural 
mechanism for $S\chi$SB due to the delocalization of single-instanton 
quark zero modes in the instanton medium, the quark condensate 
can be evaluated.  The instanton vacuum is validated by the two 
parameters: The average instanton size $\bar{\rho}\sim 1/3\, {\rm fm}$ and 
average inter-instanton distance $R\sim 1\, {\rm fm}$.  These 
essential numbers were suggested by Shuryak~\cite{Shuryak:1981ff}  
within the instanton liquid model and were derived from 
$\Lambda_{\overline{\rm MS}}$ by Diakonov and 
Petrov~\cite{Diakonov:1983hh}.  These values were recently 
confirmed by lattice QCD simulations 
\cite{Chu:vi,Negele:1998ev,DeGrand:2001tm,Faccioli:2003qz, 
Bowman:2004xi}.  
 
In the present work, we want to investigate the effect of flavor  
SU(3)-symmetry breaking on the above-mentioned three QCD condensates, 
{\it i.e.} quark, gluon, and mixed quark-gluon condensates, based on the 
instanton liquid model for the QCD vacuum~\cite{Diakonov:1985eg, 
Diakonov:1995qy,Diakonov:2002fq}.  The model was later extended by 
introducing the current quark masses~\cite{Musakhanov:1998wp, 
Musakhanov:2001pc,Musakhanov:vu,Nowak:1988bh,Alkofer:1989uj,Kacir:1996qn}.  
Since we are interested in the effect of explicit flavor 
SU(3)-symmetry breaking, we follow the formalism of 
Ref.~\cite{Musakhanov:1998wp} in which the dependence of the dynamical 
quark mass on the current quark mass has been studied in detail. 
Though the mixed quark-gluon condensate was already studied in the 
instanton vacuum~\cite{Polyakov:1996kh}, explicit SU(3)-symmetry 
breaking was not considered.  Hence, we want now to extend the work of   
Ref.~\cite{Polyakov:1996kh}, emphasizing the effect of flavor    
SU(3)-symmetry breaking on the QCD vacuum condensates.  We will show 
that with a proper choice of the $m_f$ dependence of the   
dynamical quark mass~\cite{Pobylitsa:1989uq} the gluon condensate is 
independent of $m_f$.  The corresponding results are summarized as 
follows: The ratio $[{\langle \bar{s}s\rangle}/{\langle 
\bar{u}u\rangle}]^{1/3}= 0.75$, $[{\langle 
\bar{s}\sigma_{\mu\nu}G^{\mu\nu}s\rangle}/{\langle 
\bar{u}\sigma_{\mu\nu}G^{\mu\nu}u\rangle}]^{1/5}  = 
0.87$, $m_{0,{\rm u}}^2=1.60\,{\rm GeV}^2$, and $m_{0,{\rm 
s}}^2=1.84\,{\rm GeV}^2$, with isospin symmetry assumed.      
 
\vspace{1cm} 
%==================================================================== 
{\bf 2.} We start with the effective low-energy QCD partition function 
from the instanton vacuum with SU(3)-symmetry breaking taken into 
account~\cite{Musakhanov:1998wp,Musakhanov:2001pc,Musakhanov:vu}:  
\begin{eqnarray} 
\mathcal{Z}&=&\int D\psi^{\dagger}D\psi\exp\left[\int 
  d^4x\sum_{f}{\psi}^{\dagger}_f(i\rlap{/}{\partial}+im_f)\psi_f 
\right]\left[\frac{Y_{+}^{N_f}}{VM^{N_f}}\right]^{N_{+}} 
\left[\frac{Y_{-}^{N_f}}{VM^{N_f}}\right]^{N_{-}}, 
\label{partition} 
\end{eqnarray} 
where $\psi_f$ and $\psi_f^\dagger$ denote the quark fields and $m_f$ 
stands for the current quark mass for a given flavor.  We consider 
here the strict large $N_c$ expansion~\footnote{Recently, it is  
shown that the effects of mesonic loops on the quark condensate, 
which is the next-to-leading order in the $N_c$ expansion, is not 
small~\cite{Kim:2005jc}.  However, a proper adjustment of the 
parameters $\bar{\rho}$ and $R$ may yield approximately the same results as 
those without the meson-loop corrections.}.  $V$ and $M^{N_f}$ stand 
for the space-time volume and for the dimensional parameter, 
respectively.  Note that, by dropping the current quark mass $m_f$ in 
the first bracket in the r.h.s., Eq.~(\ref{partition}) turns out to be the 
same as that derived in the chiral limit~\cite{Diakonov:1985eg, 
Diakonov:1995qy}.  $Y_{\pm}$ in Eq.~(\ref{partition}) represents 
a 't Hooft-type $2N_f$-quark interaction generated by instantons: 
\begin{eqnarray} 
Y^{N_f}_{\pm}&=&\int d\rho d(\rho)\int dU\int 
d^4x\prod_{f}\int\frac{d^4k_f}{(2\pi)^4}\frac{d^4p_f}{(2\pi)^4}[ 
2\pi\rho  F_f(k_f)][2\pi\rho F_f(p_f)]\nonumber\\&\times&\exp\left[ 
-x\cdot\left(\sum_{f}k_f-\sum_{f}p_f\right)\right]\left[ 
U^{\alpha}_{i'}(U^{j'}_{\beta})^{\dagger}\epsilon^{ii'}\epsilon_{jj'} 
\right]_f\left[i\psi_f^{\dagger}(k_f)_{\alpha i}\frac{1\pm\gamma_5}{2} 
\psi_f(p_f)^{\beta j}\right], 
\label{vertex}   
\end{eqnarray} 
where $\rho$ denotes the instanton size and $U$ represents the color 
orientation matrix. Since we are interested in the vacuum 
condensates, it is enough to consider the case of $N_f=1$ in which the 
integration over $U$ becomes trivial.  We employ the instanton 
distribution of a delta-function type, so that we get a simple 
quark-quark interaction for a given flavor $f$:  
\begin{equation} 
Y_{\pm}=\frac{i}{N_c}\int\frac{d^4k}{(2\pi)^4}[2\pi\bar{\rho}  
  F_f(k\bar{\rho})]^2\left[\psi_f^{\dagger}(k)\frac{1\pm\gamma_5}{2} 
\psi_f(k)\right].  
\label{vertex1}    
\end{equation} 
$F_f(k)$ is the Fourier transformation of the fermionic zero mode solution
$\Phi_{I\bar{I}}$. We implicitly assumed that $F_f(k)$ is a function 
of the current-quark mass. This assumption will be verified in what
follows. $\Phi_{I\bar{I}}$ satisfies the following Dirac equation under 
the 
(anti)instanton effects ${A}_{I\bar{I}}$: 
\begin{equation}
  \label{eq:zeromode}
  \left[i\rlap{/}{\partial}+\rlap{/}{A}_{I\bar{I}}+im_f\right]
\Phi_{I\bar{I}}=0.
\end{equation}
Instead of computing
$\Phi_{I\bar{I}}$ directly from Eq.~(\ref{eq:zeromode}), being equivalently, 
we follows   
the course suggested in Ref.~\cite{Pobylitsa:1989uq} in which Pobylitsa used a
elaborated and systematic expansion of the quark propagator $\langle  
x|(i\rlap{/}{\partial}+\rlap{/}{A}_I +im_f)^{-1}|x\rangle$. By doing this, one
can immediately obtain analytical form of $F_f(k)$. Here, we make a
brief explanation on this method.  Diakonov {\it et al.} made a zero-mode
approximation for the  
quark propagator in the instantons~\cite{Diakonov:1995qy}: 
\begin{equation} 
\left\langle x\left|\frac{1}{i\rlap{/}{\partial}+\rlap{/}{A}_I+im}\right|y 
\right \rangle \simeq \left\langle 
  x\left|\frac{1}{i\rlap{/}{\partial}+im}\right |y \right\rangle 
+\frac{\Phi_I(x)\Phi^{\dagger}_I(y)}{im}.    
\label{DPpro} 
\end{equation} 
We note that this zero-mode approximation causes one difficult problem of
breaking the conservation of vector and axial-vector currents. To amend this
current conservation problem, Pobylitsa expand the quark propagator as follows.
Having summed up the planar diagrams, one 
arrives the following  
integral equation for a quark propagator: 
\begin{eqnarray} 
S &=&\left\langle x\left|\frac{1}{i\rlap{/}{\partial}+\rlap{/}{A}_I+im} 
\right|x\right\rangle\nonumber\\ 
&=&\left\langle x\left|\frac{1}{i\rlap{/}{\partial}+im}\right|x 
\right\rangle + \frac{N}{2VN_c} 
{\rm tr}_{c}\left[\int d^4Z_I\left[\left\langle x\left|(i\rlap{/}{\partial} 
+\rlap{/}{A}_I+im)\right|x \right\rangle-\frac{1}{\rlap{/}{A}}\right]^{-1} 
+(I\to\bar{I})\right],   
\label{eq:pobyl} 
\end{eqnarray} 
where $Z_{I\bar{I}}$ indicates the collective coordinate for the
(anti)instanton. Then the inverse of the quark propagator can be  
approximately written with an arbitrary scalar function to be determined:   
\begin{equation} 
S^{-1} (i\partial)= i\rlap{/}{\partial}+im_f
+iM_0F^2_f(i\partial),\,\,\,\,
M_0=\frac{\lambda\left[2\pi\bar{\rho}\right]^2}{N_c}.
\label{inpro2}   
\end{equation} 
Note that we choose the scalar function to be $M_0F^2_f(i\partial)$ for later
convenience. The value of $M_0$ will be determined by the self-consistent
equation (so called ``saddle-point equation'') of the model. To obtain
$F_f(k)$, Inserting  
Eq.(\ref{inpro2}) into Eq.(\ref{eq:pobyl}), we obtain an integral equation for
$F_f(i\partial)$:   
\begin{equation} 
iF^2_f(i\partial) \simeq M_0{\rm tr}_c\left[\int 
  d^4Z_I\rlap{/}{A}_I\left(\rlap{/}{A}_I+i\rlap{/}{\partial}+im_f 
-iM_0F^2_f(i\partial)\right)^{-1}
\left(i\rlap{/}{\partial}+im_f\right) 
+(I\to\bar{I})\right]. 
\label{inpro3}   
\end{equation} 
Having carried out a straightforward manipulation, finally, we
arrive at:   
\begin{eqnarray} 
M_f(k) 
&=&M_0F^2_f(k)
\left[\sqrt{1+\frac{m^2_f}{d^2}}-\frac{m_f}{d}\right]
=M_0F^2_f(k)f(m_f),\,\,\,\,d=\sqrt{\frac{0.08385}{2N_c}}
\frac{8\pi\bar{\rho}}{R^2}   
\label{fa}   
\nonumber\\ 
F_f(k) 
&=&2t\left[I_0(t)K_1(t)-I_1(t)K_0(t)-\frac{1}{t}I_1(t)K_1(t)\right]
\simeq0.198\,{\rm  GeV},\,\,\,\,t=|k|\bar{\rho}/2.  
\label{FF1}
\end{eqnarray} 
Thus, we have exact form of $F_f(k)$ in terms of explicitly broken flavor
SU(3) symmetry ($m_f\ne0$). 

Now, we are in position to discuss the saddle-point equation of the model
derived from the effective QCD partition function. Introducing the Lagrange
multipliers  
$\lambda_\pm $~\cite{Diakonov:2002fq}, we are able to exponentiate 
the partition function of Eq.~(\ref{partition}): 
\begin{eqnarray} 
\mathcal{Z}&=& 
\int\frac{d\lambda_{\pm}}{2\pi}\int D\psi^{\dagger}D\psi\exp 
\left[\int d^4x\bar{\psi}_f(i\rlap{/}{\partial}+im_f)\psi_f\right]\nonumber\\ 
&\times&\exp\left[N_+\left(\ln\frac{N_+}{\lambda_+VM^{N_f}}-1\right) 
+\lambda_+Y_{+}\right]\exp\left[N_-\left(\ln\frac{N_-}{\lambda_- 
VM^{N_f}}-1\right)+\lambda_-Y_{-}\right].   
\end{eqnarray} 
In the following, we assume that $N_+=N_-=N/2$ and 
$\lambda_+=\lambda_-=\lambda$ so that the partition function can be 
simplified as follows:  
\begin{equation} 
\mathcal{Z}=\int\frac{d\lambda}{2\pi}\int D\psi^{\dagger} 
D\psi\exp\left[\int d^4x\psi_f^{\dagger}(i\rlap{/}{\partial}+im_f) 
\psi_f+N\left(\ln\frac{N}{2\lambda VM^{N_f}}-1\right)+ 
\lambda(Y_{+}+Y_{-})\right].\nonumber\\ 
\label{partion3}   
\end{equation} 
 
It is also interesting to compare Eq.~(\ref{partion3}) with the 
partition function proposed by Diakonov {\it et al.}: 
\begin{equation} 
\mathcal{Z}_{D}=\int\frac{d\lambda}{2\pi}\int 
D\psi^{\dagger}D\psi\exp\left[\int 
d^4x\psi_f^{\dagger}i\rlap{/}{\partial}\psi_f+N\left(\ln\frac{N}{ 
2\lambda VM^{N_f}}-1\right)+\lambda(Y_{+}+Y_{-}+2m_f)\right]. 
\label{partion4}   
\end{equation} 
One can easily find the difference between the effective actions of 
Eq.~(\ref{partion3}) and of Eq.~(\ref{partion4}); In 
Eq.~(\ref{partion3}), the current quark mass $m_f$ appears as a mass 
term of the quark fields accounting for the explicit breaking of 
chiral symmetry, while the $m_f$ is placed in the quark-instanton 
vertex in Eq.~(\ref{partion4}).  We note that this configuration    
of the current quark mass is deeply related to the parity-violating 
quark mass term ($\propto \psi^{\dagger}\gamma_5\psi$) in terms of the 
instanton number fluctuation~\cite{Diakonov:1995qy}.     
 
Concentrating on the first and third brackets in Eq.~(\ref{partion3}), 
we get the fermionic trace log which relates to the quark propagator 
as follows:  
\begin{eqnarray} 
&&\int d^4x\int\frac{d^4k}{(2\pi)^4}{\rm tr}_{c\gamma}\ln\frac{\left[ 
-\rlap{/}{k}+im_f+\frac{i\lambda}{N_c}[2\pi\bar{\rho}F_f(k\bar{\rho})]^2\right]} 
{\left[-\rlap{/}{k}+im_f\right]}\nonumber\\&&\hspace{2cm}=N_cV\int 
\frac{d^4k}{(2\pi)^4}{\rm tr}_{\gamma}\ln\frac{\left[-\rlap{/}{k}+im_f 
+iM_f(k)\right]}{ 
\left[-\rlap{/}{k}+im_f\right]},    
\end{eqnarray} 
where the subscripts $c$ and $\gamma$ denote the color and Dirac-spin   
spaces.  Thus, we obtain the partition function in a compact form as   
follows: 
\begin{equation} 
\mathcal{Z}=\int\frac{d\lambda}{2\pi}\exp\left[N\left(\ln\frac{N}{2\lambda 
VM^{N_f}}-1\right)+N_cV\int\frac{d^4k}{(2\pi)^4}{\rm 
tr}_{d}\ln\frac{\left[-\rlap{/}{k}+im_f+iM_f(k)\right]}
{\left[-\rlap{/}{k}+im_f\right]}\right].\nonumber\\    
\label{effectiveZ}  
\end{equation}  
Now, we perform a functional variation of the partition function with 
respect to $\lambda$: 
\begin{eqnarray} 
\frac{\delta\ln\mathcal{Z}}{\delta\lambda}&=&-\frac{N}{\lambda}+ 
N_cV\int\frac{d^4k}{(2\pi)^4}{\rm tr}_\gamma\frac{{iM_f(k)}/{N_c}}
{-\rlap{/}{k}+im_f+iM_f(k)}=0,\nonumber\\  
\frac{N}{\lambda}&=&N_cV\int\frac{d^4k}{(2\pi)^4}{\rm tr}_\gamma 
\frac{iM_f(k) [-\rlap{/}{k}-im_f-iM_f(k)]}{k^2+[m_f+M_f(k])^2},
\nonumber\\\frac{N}{V}&=&4N_c\int\frac{d^4k}{ 
(2\pi)^4}\frac{M_f(k)[m_f+M_f(k)]}{k^2+[m_f+M_f(k)]^2}. 
\label{SPEq}   
\end{eqnarray}
Final formula in Eq.~(\ref{SPEq}) is called the saddle-point equation of the
model. We employ the  
standard values of the instanton ensemble, $N/V=200^4$ MeV$^4$ and 
$\bar{\rho}\sim1/3$ fm $\simeq1/600$ MeV$^{-1}$ for the numerical 
calculations. From these values, we obtain $M_0=0.350$ GeV, which is
determined self-consistently by the saddle-point equation. Using this value,
we obtain $i\langle u^{\dagger}u\rangle\simeq250^3$ MeV$^3$ and pion weak
decay constant $F_{\pi}\simeq93$ MeV. We note that this saddle-point equation
is closely related to the  
gluon-condensate~\cite{Schafer:1996wv}; $\langle  
G_{\mu\nu}G^{\mu\nu}\rangle_f=32\pi^2N/V$.  
 
As for an arbitrary $N_f$, it becomes rather difficult to obtain the 
saddle-point equation, since we need to perform a complicated 
integration over the color orientation.  However, keeping only the 
leading order in the large $N_c$ limit and additional variation over 
the Hermitian $N_f\times N_f$ flavor matrix ($\mathcal{M}$ in 
Ref.~\cite{Diakonov:1995qy}), we have the same form of the 
saddle-point equation for each flavor as shown in 
Eq.~(\ref{SPEq}). The similar argument is also possible for the 
quark and mixed condensates~\cite{Musakhanov:2001pc,Polyakov:1996kh, 
Diakonov:1995qy}.  
 
Being similar to the saddle-point equation (the gluon condensate), the 
quark condensate for each flavor can be obtained by the following 
functional derivative with respect to the current quark mass $m_f$: 
\begin{eqnarray} 
\langle\bar{q}q\rangle_f&=&\frac{1}{V}\frac{\delta\ln\mathcal{Z}}{ 
\delta m_f}=-iN_c\int\frac{d^4k}{(2\pi)^4}{\rm tr}_{\gamma}\left[\frac{ 
\rlap{/}{k}+i[m_f+M_f(k)]}{k^2+[m_f+M_f(k)]^2}-\frac{\rlap{/}{k}+im_f}{ 
k^2+m_f^2}\right]\nonumber\\&=&4N_c\int\frac{d^4k}{(2\pi)^4}\left[ 
\frac{m_f+M_f(k)}{k^2+[m_f+M_f(k)]^2}-\frac{m_f}{k^2+m_f^2}\right]. 
\label{qq}   
\end{eqnarray} 
One can immediately see that when $m_f\to0$, Eq.~(\ref{qq}) gets equal  
to the well-known expression for the quark condensate in the chiral 
limit.  
 
Now, we discuss how to calculate the mixed condensate,  
$\langle\bar{q}\sigma_{\mu\nu}G^{\mu\nu}q\rangle$.  Actually, the local 
operator inside this condensate corresponds to the quark-gluon 
interaction of a Yukawa type.  However, in the present work, the gluon field 
strength ($G_{\mu\nu}$) can be expressed in terms of the quark-instanton 
interaction~\cite{Diakonov:1995qy,Polyakov:1996kh}.  First, the one 
flavor quark and one instanton interaction in Eq.~(\ref{vertex}) can be 
rewritten as a function of space-time coordinates $x$ and color 
orientation matrix $U$:    
\begin{eqnarray} 
Y_{\pm,1}(x,U)&=&\int\frac{d^4k_1}{(2\pi)^4}\frac{d^4k_2}{(2\pi)^4}[ 
2\pi\rho F_f(k_1)][2\pi\rho F_f(k_2)]e^{-ix\cdot(k_1-k_2)}\nonumber\\ 
&\times&\left[U^{\alpha}_{i'}(U^{j'}_{\beta})^{\dagger}\epsilon^{ii'} 
\epsilon_{jj'}\right]_f\left[i\psi_f^{\dagger}(k_1)_{\alpha i} 
\frac{1\pm\gamma_5}{2}\psi_f(k_2)^{\beta j}\right]. 
\label{vertex2}   
\end{eqnarray} 
Here, we assume again the delta function-type instanton distribution. We 
define then the field strength $G^a_{\mu\nu}$ in terms of the instanton 
configuration as a function of $I$($\bar{I}$) position and orientation 
matrix $U$:  
\begin{equation} 
G^a_{\pm\mu\nu}(x,x',U)=\frac{1}{2}\left[\lambda^aU\lambda^b 
U^{\dagger}\right]G^b_{\pm\mu\nu}(x'-x).  
\label{field}   
\end{equation} 
$G^b_{\pm\mu\nu}(x'-x)$ stands for the field strength consisting of a 
certain instanton configuration.  Using Eqs.~(\ref{vertex2}) and 
(\ref{field}), we define the field strength in terms of the  
quark-instanton interaction:  
\begin{equation} 
\hat{G}^a_{\pm\mu\nu}=\frac{iN_cM}{4\pi\bar{\rho}^2}\int d^4x 
\int dU G^a_{\pm\mu\nu}(x,x',U)Y_{\pm,1}(x,U)    
\end{equation} 
Following the method in Ref.~\cite{Diakonov:1995qy,Polyakov:1996kh}, 
we finally obtain the quark-gluon mixed condensate as follows: 
\begin{equation} 
\langle\bar{q}\sigma_{\mu\nu}G^{\mu\nu}q\rangle_f=2N_c\bar{\rho}^2\int 
\frac{d^4k_1}{(2\pi)^4}\int\frac{d^4k_2}{(2\pi)^4}\frac{ 
\sqrt{M_f(k_1)M_f(k_2)}G(k_1,k_2)N(k_1,k_2)}{[k^2_1+[m_f+M_f(k_1)]^2][ 
k^2_2+[m_f+M_f(k_2)]^2]},\nonumber\\  
\label{mc}   
\end{equation} 
where $G(k_1,k_2)$ and $N(k_1,k_2)$ are defined as follows: 
\begin{eqnarray} 
G(k_1,k_2)&=&32\pi^2\left[\frac{K_0(t)}{2}+\left[\frac{4K_0(t)}{t^2} 
+\left(\frac{2}{t}+\frac{8}{t^3}\right)K_1(t)-\frac{8}{t^4}\right]\right], 
\nonumber\\  
N(k_1,k_2)&=&\frac{1}{(k_1-k_2)^2}\left[8k^2_1k^2_2-6(k^2_1+k^2_2)k_1\cdot 
  k_2+4(k_1\cdot k_2)^2\right]   
\end{eqnarray} 
with $t=|k_1-k_2|\bar{\rho}$. The functions $K_0$ and $K_1$ stand for the 
modified Bessel functions of the second kind of order $0$ and $1$, 
respectively. If we consider for arbitrary $N_f$ the mixed condensate, 
the situation may be somewhat different from the cases of the gluon 
and quark condensates. We note that, though the mixed condensate has 
been calculated for the case of $N_f=1$, the same formula of 
Eq.~(\ref{mc}) still holds for each flavor with arbitrary $N_f$ as 
discussed previously~\cite{Polyakov:1996kh}.    
 
\begin{figure}[t] 
\begin{tabular}{cc} 
\includegraphics[width=10cm]{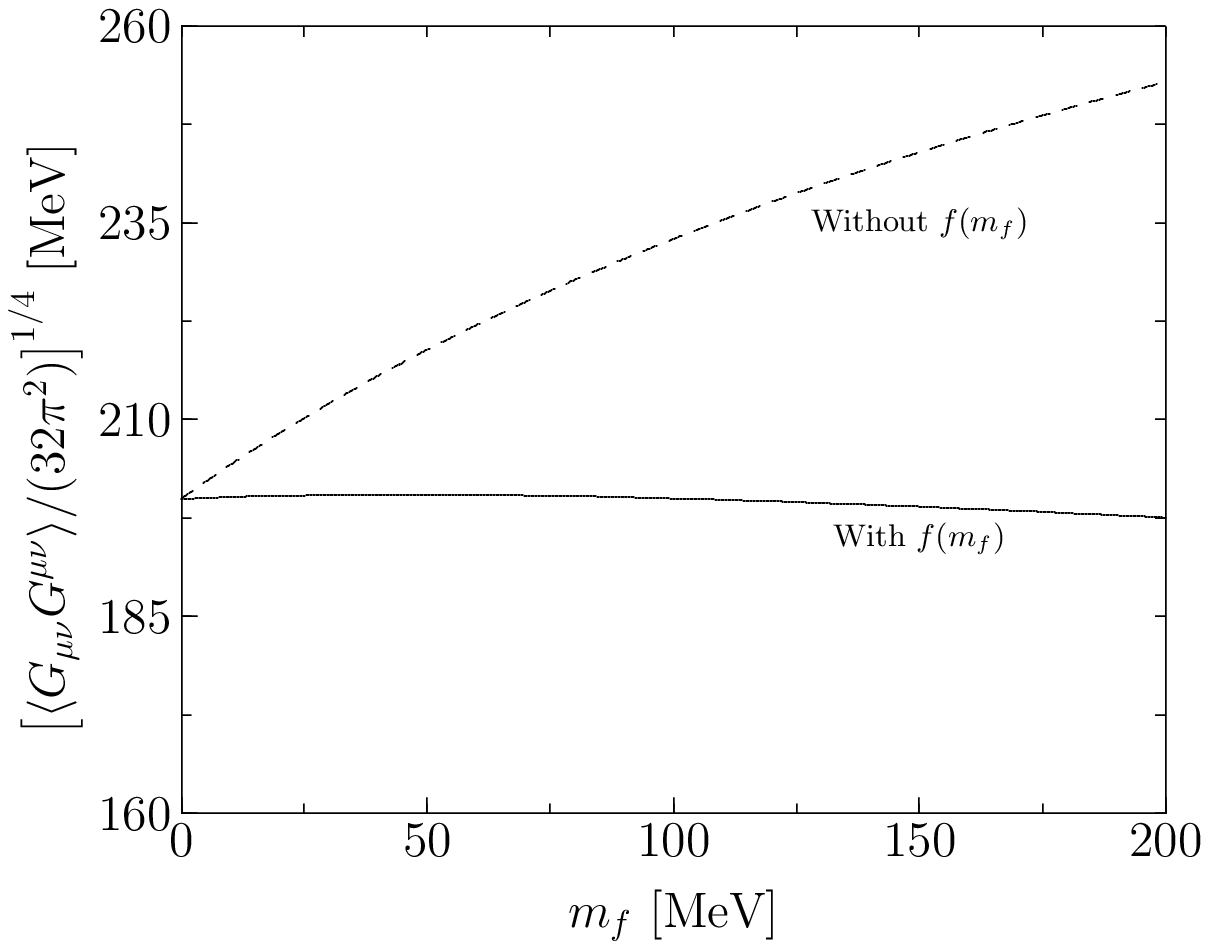} 
\end{tabular} 
\caption{Gluon condensate as a function of the current quark mass   
$m_f$.  The solid curve draws the gluon condensate with the 
$m_f$-correction factor $f(m_f)$ in Eq.~(\ref{fa}) and the dashed one 
corresponds to that without it.}     
\label{fig2}  
\end{figure}

\vspace{1cm} 
%=======================================================================  
{\bf 3.} We first discuss the numerical results for the  
gluon condensate, $\langle G_{\mu\nu}G^{\mu\nu}\rangle/32\pi^2=N/V$. 
As mentioned before, we use $N/V= 200^4$ MeV$^4$~\cite{Schafer:1996wv}.   
In Fig.~\ref{fig2}, we show the  numerical results of the gluon condensates 
with and without the $m_f$-correction factor $f(m_f)$ in 
Eq.~(\ref{fa}).  As shown in Fig.~\ref{fig2}, the result with $f(m_f)$ 
is noticeably different from that without it.  Since the gluon 
condensate is related to the vacuum energy density, it should be 
independent of the current quark mass $m_f$.  However, if we turn 
off the $m_f$-correction factor, the gluon condensate increases almost 
linearly.  It indicates that it is essential to have the 
$m_f$-correction factor $f(m_f)$ in order to consider SU(3) flavor 
symmetry breaking properly.  In Table~\ref{table1}   
we list the values of the gluon condensate when  
$m_f=m_u=5$ MeV and $m_f=m_s=150$ MeV with and without $f(m_f)$.      
 
\begin{table}[b] 
\begin{tabular}{c|c|c} 
\hline 
\hline 
&With $f(m_f)$&without $f(m_f)$\\ 
\hline 
$\langle G_{\mu\nu}G^{\mu\nu}\rangle_u/32\pi^2$ &$200^4$ &$202^4$ \\  
$\langle 
G_{\mu\nu}G^{\mu\nu}\rangle_s/32\pi^2$ & $199^4$ &$243^4$\\ 
\hline 
\hline 
\end{tabular} 
\caption{Gluon condensates for $m_u=5$ and $m_s=150$ MeV [MeV$^4$].}  
\label{table1} 
\end{table} 
 
\begin{figure}[t] 
\begin{tabular}{cc} 
\includegraphics[width=10cm]{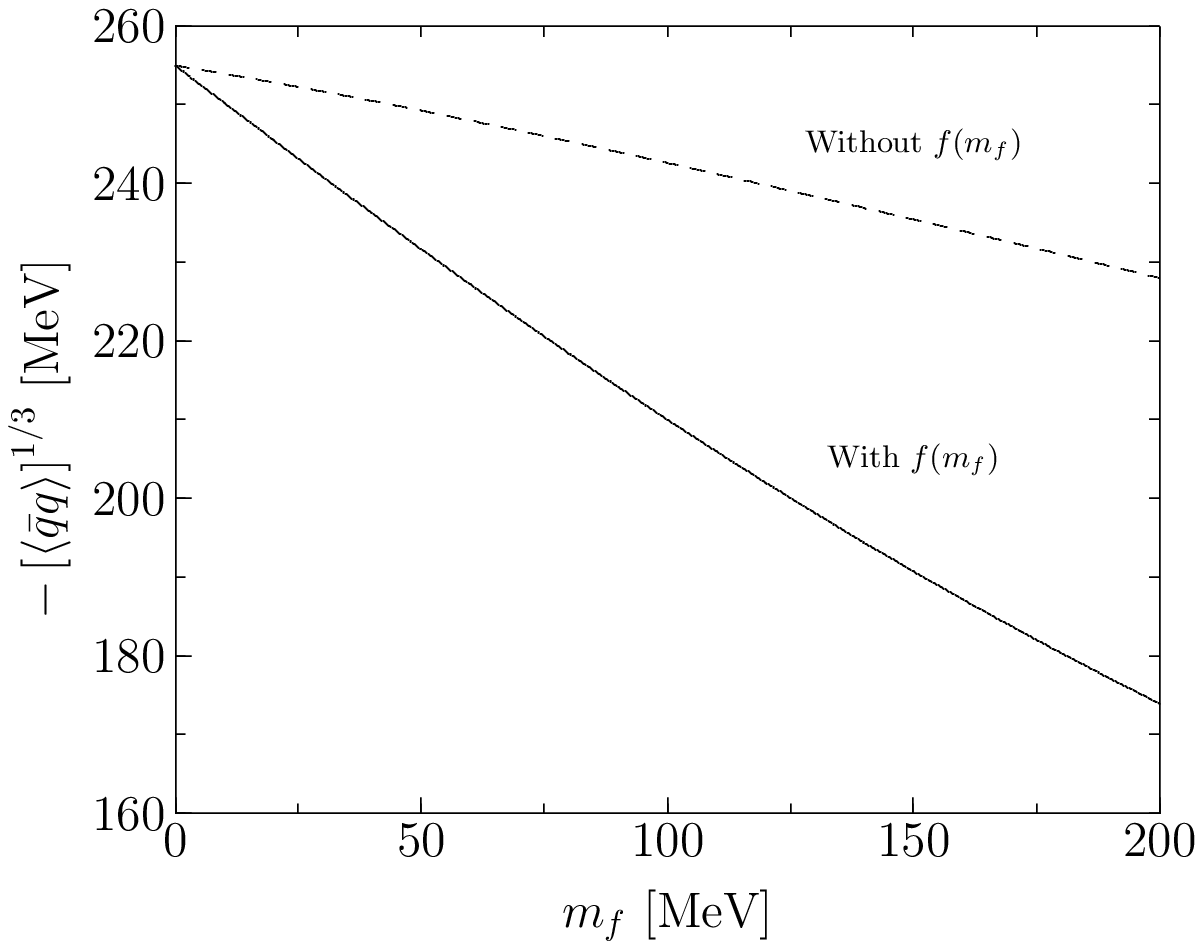} 
\end{tabular} 
\caption{Quark condensate as a function of the current quark mass   
$m_f$.  The solid curve draws the quark condensate with the 
$m_f$-correction factor $f(m_f)$ in Eq.~(\ref{fa}) and the dashed one 
corresponds to that without it.} 
\label{fig3}  
\end{figure}

In Fig.~\ref{fig3}, we depict the results of the quark condensate in a 
similar manner.  In general, the quark condensate decreases as the 
value of $m_f$ increases, as already shown in Ref.~\cite{Kim:2005jc}. 
The values of the quark condensate are listed in Table~\ref{table2}.   
As for the $u$-quark condensate, we have $-250^3\sim-260^3$ MeV$^3$. 
Here, we assume isospin symmetry for the light quarks; $m_u=m_d=5$ 
MeV.  We find that the nonstrange quark condensate does not depend  
much on $f(m_f)$.  In contrast, the strange one is sensitive to the 
$m_f$ correction factor as shown in Table~\ref{table2} (and also in  
Fig.~\ref{fig3}).  We find that without the $f(m_f)$ the quark 
condensate is decreased by about 10 \% as $m_f$ increases from 0 to 
200 MeV, while it is diminished by about 30 \% with the $f(m_f)$. 
Being compared with the results of 
Refs.~\cite{Musakhanov:1998wp,Jamin:2002ev,Dosch:1988vv},  
the quark condensate in the present work is obtained to be similar to 
them with the correction factor. 
\begin{table}[b] 
\begin{tabular}{c|c|c} 
\hline 
\hline 
&With $f(m_f)$&without $f(m_f)$\\ 
\hline 
$\langle\bar{u}u\rangle$&$-$253$^3$&$-$254$^3$ 
\\$\langle\bar{s}s\rangle$&$-$191$^3$&$-$235$^3$\\ 
\hline 
\hline 
\end{tabular} 
\caption{Quark condensates for $m_u=5$ and $m_s=150$ MeV [MeV$^3$].} 
\label{table2} 
\end{table} 
 
\begin{figure}[t] 
\begin{tabular}{cc} 
\includegraphics[width=10cm]{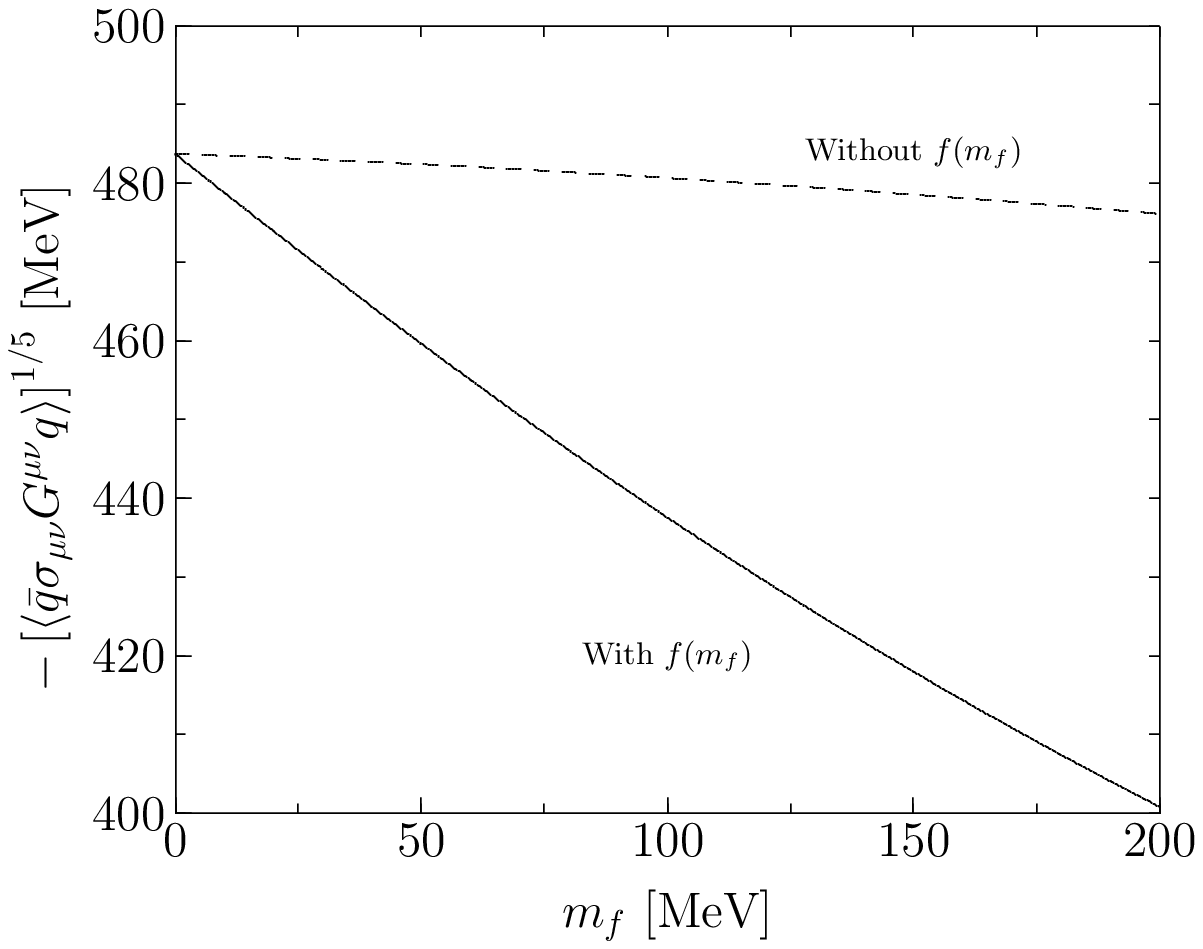} 
\end{tabular} 
\caption{Quark-gluon mixed condensate as a function of the current 
quark mass $m_f$.  The solid curve draws the mixed condensate with the 
$m_f$-correction factor $f(m_f)$ in Eq.~(\ref{fa}) and the dashed one 
corresponds to that without it.}  
\label{fig4}  
\end{figure}

In Fig.~\ref{fig4}, we draw the results of the quark-gluon mixed 
condensate as functions of the $m_f$.  The correction factor $f(m_f)$ 
being taken into account, the mixed condensate is decreased by about 
15 \%.  However, it is almost constant when $f(m_f)$ is switched off.   
We also list the values of the mixed condensate for the up and  
strange quarks in Table~\ref{table3}.  
\begin{table}[b] 
\begin{tabular}{c|c|c} 
\hline 
\hline 
&With $f(m_f)$&without $f(m_f)$\\ 
\hline 
$\langle\bar{u}\sigma_{\mu\nu}G^{\mu\nu}u\rangle$&$-$481$^5$&$-$484$^5$\\ 
$\langle\bar{s}\sigma_{\mu\nu}G^{\mu\nu}s\rangle$&$-$418$^5$&$-$483$^5$\\ 
\hline 
\hline 
\end{tabular} 
\caption{Quark-gluon mixed condensates for $m_u=5$ and $m_s=150$ MeV 
  [MeV$^5$].}  
\label{table3} 
\end{table} 
 
We now consider the effect of flavor SU(3)-symmetry breaking by 
calculating the ratios between the nonstrange condensates and the 
strange ones.  As already discussed, the ratio of the gluon 
condensates remains unity for the $m_f$-correction factor.  As 
for the ratio of the quark condensate, $[\langle\bar{s}s\rangle/\langle 
\bar{u}u\rangle]^{1/3}$, there is a great amount of theoretical 
calculations and those results lie in the range: 
$0.79\sim1.03$~\cite{Musakhanov:1998wp,Jamin:2002ev,Dosch:1988vv}.  Our 
results are $0.75\sim 0.93$ for the quark  
condensate, which is consistent with them.  We note that 
$[\langle\bar{s}s\rangle/\langle \bar{u}u\rangle]^{1/3}=0.75$ with 
the $m_f$-correction factor will be taken as our best result.  Being 
compared to Refs.~\cite{Beneke:1992ba,Aladashvili:1995zj, 
Khatsimovsky:1987bb,Ovchinnikov:1988gk}, our results are in good 
agreement with them.     
  
\begin{figure}[t] 
\begin{tabular}{cc} 
\includegraphics[width=10cm]{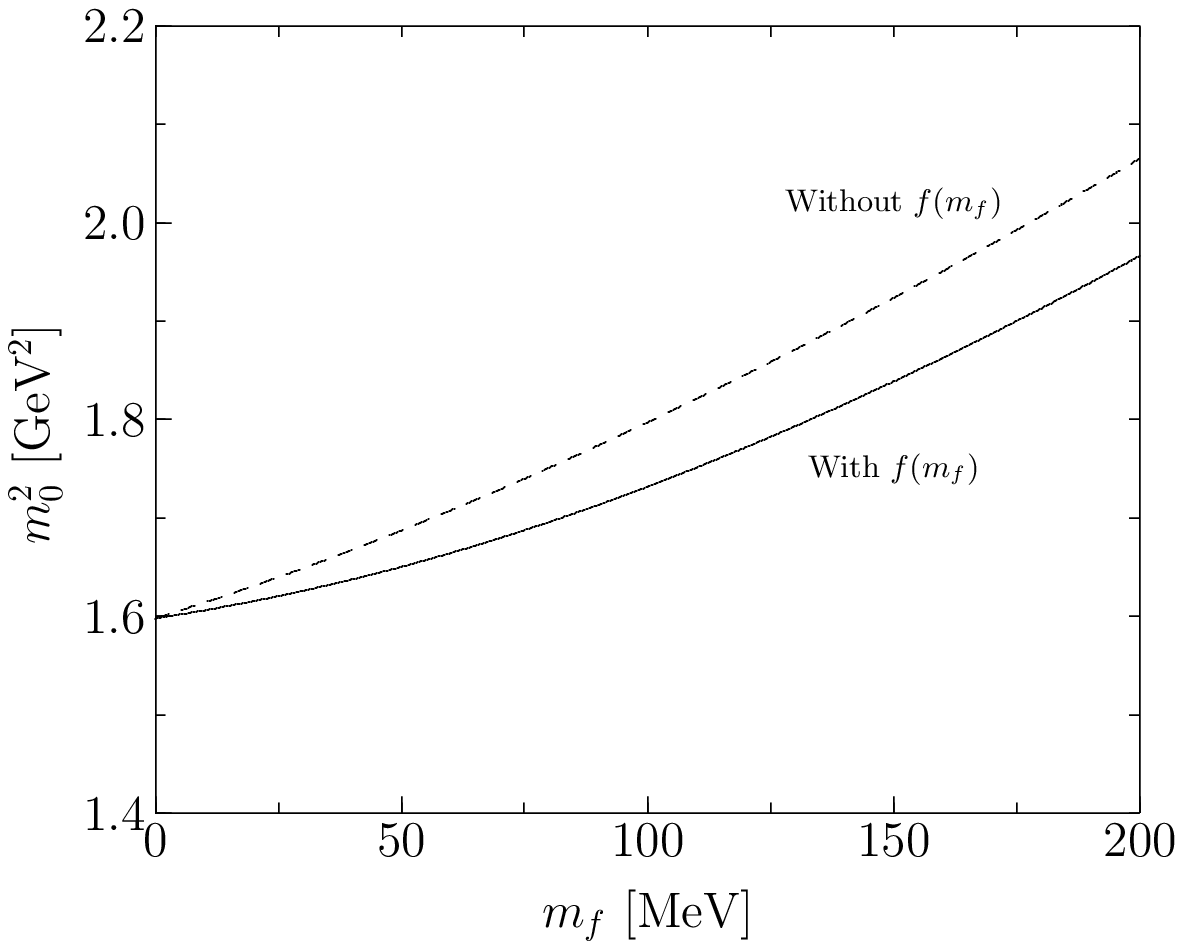} 
\end{tabular} 
\caption{Ratio $m^2_0$ as a function of the current quark mass   
$m_f$.  The solid curve draws the gluon condensate with the 
$m_f$-correction factor $f(m_f)$ in Eq.~(\ref{fa}) and the dashed one 
corresponds to that without it.}  
\label{fig5}  
\end{figure}

We now study a dimensional quantity $m_0^2$ defined as the ratio 
between the mixed and quark condensates:  
\begin{eqnarray}                 
m^2_0=\langle\bar{q}\sigma_{\mu\nu}G^{\mu\nu}q\rangle/\langle\bar{q}q\rangle, 
\end{eqnarray}               
which is an important input for general QCD sum rule calculations. We 
draw $m_0^2$ in Fig.~\ref{fig5} as a function of $m_f$ 
and list the values of $m^2_0$ in Table~\ref{table4}. 
\begin{table}[b] 
\begin{tabular}{c|c|c|c} 
\hline 
\hline 
&With $f(m_f)$&without $f(m_f)$&Other results\\ 
\hline 
$[\langle G_{\mu\nu}G^{\mu\nu}\rangle_s/\langle 
G_{\mu\nu}G^{\mu\nu}\rangle_u]^{1/4}$&1.00&1.20&$-$\\  
$[\langle\bar{s}s\rangle/\langle 
\bar{u}u\rangle]^{1/3}$&0.75&0.93&$0.79\sim1.03$ 
~\cite{Musakhanov:1998wp,Jamin:2002ev,Dosch:1988vv}\\   
$[\langle\bar{s}\sigma_{\mu\nu}G^{\mu\nu}s\rangle/\langle\bar{u} 
\sigma_{\mu\nu}G^{\mu\nu}u\rangle]^{1/5} $&0.87&1.00& 
$0.75\sim1.05$~\cite{Beneke:1992ba,Aladashvili:1995zj,Khatsimovsky:1987bb, 
Ovchinnikov:1988gk,Braun:2004vf}\\  
$m^2_{0,u}=\langle\bar{u}\sigma_{\mu\nu}G^{\mu\nu}u\rangle/ 
\langle\bar{u}u\rangle$&1.60 GeV$^2$&1.60 
GeV$^2$&$0.8\pm0.2$~\cite{Belyaev:1982sa}, 
$1.4$~\cite{Polyakov:1996kh}, $2.5$~\cite{Doi:2002wk} [GeV$^2$]\\  
$m^2_{0,s}=\langle\bar{s}\sigma_{\mu\nu}G^{\mu\nu}s\rangle/ 
\langle\bar{s}s\rangle$&1.84 GeV$^2$&1.92 GeV$^2$&$-$\\ 
\hline 
\hline 
\end{tabular} 
\caption{The ratios of the condensates for the different types of the 
  $m_f$ correction factors. $m_u=5$ MeV and $m_s=150$ MeV are used.} 
\label{table4} 
\end{table}  
The value of $m_0^2$ increases as $m_f$ does, which implies that the 
mixed condensate is less sensitive to the $m_f$ than the quark 
condensate.  The values of $m_0^2$ are in the range of $1.84\, {\rm 
  GeV}^2$ for the strange quark and of $1.60 \,{\rm GeV}^2$ for the up 
quark.  Thus, the strange $m_{0,s}^2$ turns out to be larger than the 
nonstrange $m_{0,u}^2$ by about $15\,\%$. 
 
We now consider the scale dependence of the condensates. 
\begin{equation} 
\langle\mathcal{O}_1(\mu_A)\rangle=\left[\frac{\alpha_s(\mu_B)} 
{\alpha_s(\mu_A)}\right]^{\gamma_1/b}\langle\mathcal{O}_1(\mu_B) 
\rangle,\,\,\,\langle\mathcal{O}_2(\mu_A)\rangle=\left[ 
\frac{\alpha_s(\mu_B)}{\alpha_s(\mu_A)}\right]^{\gamma_2/b} 
\langle\mathcal{O}_2(\mu_B)\rangle,  
\label{norm}   
\end{equation} 
where the subscripts $1$ and $2$ indicate different generic operators 
corresponding to the condensates in which we are interested.  The 
$\mu_A$ and $\mu_B$ denote two different renormalization scales of the 
operators.  $\gamma$ is the corresponding anomalous dimension for 
the condensates.  $b$ is defined as $11N_c/3-2N_f/3$ and becomes 
$11-2=9$ in the present work.  Taking the ratio of the operators $1$ 
and $2$, we have:  
\begin{equation} 
\frac{\langle\mathcal{O}_1(\mu_A)\rangle}{\langle\mathcal{O}_2(\mu_A) 
\rangle}=\left[\frac{\alpha_s(\mu_B)}{\alpha_s(\mu_A)}\right]^{ 
(\gamma_1-\gamma_2)/b}\frac{\langle\mathcal{O}_1(\mu_B)\rangle}{ 
\langle\mathcal{O}_2(\mu_B)\rangle}. 
\label{norm1}   
\end{equation} 
If $\mathcal{O}_1$ and $\mathcal{O}_2$ stand for the mixed and 
quark condensates respectively, we have $\gamma_1=-2/3$ and 
$\gamma_2=4$ resulting in $(\gamma_1-\gamma_2)/b=-14/27 \simeq-0.52$.  On 
the contrary, we have $\gamma_1/b=-2/27\simeq-0.07$  
and $\gamma_2/b=4/9\simeq 0.44$.  Thus, $m^2_0$ shows relatively 
strong dependence on the scale, whereas the quark and 
mixed condensates are less influenced by the scaling.  Thus, $m_0^2$ 
depends more strongly on the scale.   
 
Finally, we would like to mention that the relation between the mixed 
condensate and the expectation value of the transverse momentum for 
the leading-twist light-cone distribution amplitude.  This relation 
can be written as follows:  
\begin{equation} 
\langle k^2_T\rangle_{\pi}=\frac{5}{36}\frac{\langle\bar{u} 
\sigma_{\mu\nu}G^{\mu\nu}u\rangle}{\langle 
\bar{u}u\rangle}=\frac{5m^2_0}{36}.   
\end{equation} 
Considering $m^2_0$ for the light quarks in Table~\ref{table4} 
($m^2_0=0.16\, {\rm GeV}^2$), we get $\langle k^2_T\rangle_{\pi}= 
0.2\,{\rm GeV}^2$.  Note that the value estimated here is in  
qualitatively good agreement with that calculated directly from the 
pseudoscalar meson light-cone distribution amplitude within the same 
framework~\cite{Nam:2006au}; $\langle k^2_T\rangle_{\pi}=0.23\,{\rm 
GeV}^2$.   
 
\vspace{1cm}  
{\bf 4.} In the present work, we investigated the various QCD vacuum 
condensates within the framework of the instanton liquid model, 
emphasizing the effects of flavor SU(3)-symmetry breaking.  The 
modified improved action elaborated by Musakhanov was used for 
this purpose.  In the modified improved action, the current  
quark mass appeared explicitly in the denominator of the quark 
propagator as well as in the dynamical quark mass.  Thus, we were able 
to take into account the current quark mass effects to the QCD  
condensates.  In order to consider the $m_s$ dependence of the 
dynamical quark mass, we employed the $m_f$-correction factor. It arises 
from the resummation of the QCD planar loops in the large $N_c$ limit 
and is parameterized to satisfy the saddle-point 
equation~\cite{Pobylitsa:1989uq,Musakhanov:2001pc}.  
 
We observed that the gluon condensate is almost independent of the 
current quark mass when the strange current quark correction factor is 
introduced, whereas it increases monotonically without it.  The quark 
and mixed condensates were also calculated.  The results were 
consistent with those from other model calculations as well as 
phenomenological values.  In particular,  The ratios of the 
condensates between the strange and up quarks were also investigated:  
$[\langle\bar{s}\sigma_{\mu\nu}G^{\mu\nu}s\rangle/\langle\bar{u} 
\sigma_{\mu\nu}G^{\mu\nu}u\rangle]^{1/5}$ and 
$[\langle\bar{s}s\rangle/ \langle \bar{u}u\rangle]^{1/3}$.  It turned 
out that the results are again compatible with other theoretical 
calculations.  The dimensional quantity $m^2_{0}$ was also studied: 
$m^2_{0,u}=1.60\,{\rm GeV}^2$ and $m^2_{0,s}=1.84\,{\rm GeV}^2$.  In 
general, the quark and mixed condensates decrease as the current   
quark mass increases.  However, the $m_0^2$ increases as the current 
quark mass does, which indicates that the mixed condensate is less 
sensitive to the current quark mass than the quark condensate.     
In addition, we also presented the 
relation of the $m_0^2$ to the expectation value of the transverse 
momentum for the leading-twist pion light-cone distribution 
amplitude. 
 
Though we considered the effects of flavor SU(3)-symmetry breaking on 
various condensates, we did not take into account those from 
meson-loop corrections which are known to be of importance.  The 
corresponding investigation is under way.   
 
\section*{Acknowledgments} 
The present work is supported by the Korea Research Foundation Grant
funded by the Korean Government(MOEHRD) (KRF-2006-312-C00507).  The
work of S.N. is supported by the Brain Korea 21 (BK21) project in
Center of Excellency for Developing Physics Researchers of Pusan
National University, Korea.  The authors are grateful to
M.~M.~Musakhanov for critical comments and discussions.   
%================================================= 
 

\begin{thebibliography}{99} 
%================================================= 
\bibitem{Shifman:1978bx} 
M.~A.~Shifman, A.~I.~Vainshtein and V.~I.~Zakharov, 
Nucl.\ Phys.\ B {\bf 147} (1979)  385; Nucl.\ Phys.\ B {\bf 147} (1979) 448. 
%================================================= 
\bibitem{Gasser:1983yg} 
J.~Gasser and H.~Leutwyler, Annals Phys.\  {\bf 158} (1984) 142. 
%================================================= 
\bibitem{Gell-Mann:1968rz} 
M.~Gell-Mann, R.~J.~Oakes and B.~Renner, Phys.\ Rev.\  {\bf 175} (1968) 2195. 
%================================================= 
\bibitem{Shifman:1978by} 
M.~A.~Shifman, A.~I.~Vainshtein and V.~I.~Zakharov, 
Nucl.\ Phys.\ B {\bf 147} (1979) 448. 
%================================================= 
\bibitem{Ioffe:1981kw} 
B.~L.~Ioffe, Nucl.\ Phys.\ B {\bf 188} (1981) 317 
[Erratum-ibid.\ B {\bf 191} (1981) 591]. 
%================================================= 
\bibitem{Balitsky:1986hf} 
I.~I.~Balitsky, D.~Diakonov and A.~V.~Yung,Z.\ Phys.\ C {\bf 33} (1986) 265. 
%================================================= 
\bibitem{Chernyak:1983ej} 
V.~L.~Chernyak and A.~R.~Zhitnitsky, Phys.\ Rept.\  {\bf 112} (1984) 173. 
%================================================= 
\bibitem{Belyaev:1982sa} 
V.~M.~Belyaev and B.~L.~Ioffe, Sov.\ Phys.\ JETP {\bf 56} (1982) 493 [Zh.\ 
Eksp.\ Teor.\ Fiz.\  {\bf 83} (1982) 876]. 
%================================================= 
\bibitem{Dosch:1988vv} 
H.~G.~Dosch, M.~Jamin and S.~Narison, Phys.\ Lett.\ B {\bf 220} (1989) 251. 
%================================================= 
\bibitem{Dorokhov} 
A.~E.~Dorokhov, S.~V.~Esaibegian and S.~V.~Mikhailov, 
  %``Virtualities of quarks and gluons in QCD vacuum and nonlocal  condensates 
  %within single instanton approximation,'' 
  Phys.\ Rev.\ D {\bf 56} (1997) 4062. 
%================================================ 
\bibitem{Polyakov:1996kh} 
M~.V~.~Polyakov and C.~Weiss, Phys.\ Lett.\ B {\bf 387} (1996) 841.  
%================================================= 
\bibitem{Doi:2002wk} 
T.~Doi, N.~Ishii, M.~Oka and H.~Suganuma, Phys.\ Rev.\ D {\bf 67} 
(2003) 054504. 
%================================================= 
\bibitem{Shuryak:1981ff} 
E.~V.~Shuryak, Nucl.\ Phys.\ B {\bf 203} (1982) 93. 
%================================================= 
\bibitem{Diakonov:1983hh} 
D.~Diakonov and V.~Y.~Petrov, Nucl.\ Phys.\ B {\bf 245} (1984) 259. 
%================================================= 
\bibitem{Chu:vi} 
M.~C.~Chu, J.~M.~Grandy, S.~Huang and J.~W.~Negele, Phys.\ Rev.\ D {\bf 49} 
 (1994) 6039.  
%================================================= 
\bibitem{Negele:1998ev} 
J.~W.~Negele, Nucl.\ Phys.\ Proc.\ Suppl.\  {\bf 73} (1999) 92. 
%================================================= 
\bibitem{DeGrand:2001tm} 
T.~DeGrand, Phys.\ Rev.\ D {\bf 64} (2001) 094508. 
%================================================= 
\bibitem{Faccioli:2003qz} 
P.~Faccioli and T.~A.~DeGrand,Phys.\ Rev.\ Lett.\  {\bf 91} (2003) 182001. 
%================================================= 
\bibitem{Bowman:2004xi} 
P.~O.~Bowman, U.~M.~Heller, D.~B.~Leinweber, A.~G.~Williams and 
J.~b.~Zhang, Nucl.\ Phys.\ Proc.\ Suppl.\  {\bf 128} (2004) 23. 
%================================================= 
\bibitem{Diakonov:1985eg} 
D.~Diakonov and V.~Y.~Petrov,Nucl.\ Phys.\ B {\bf 272} (1986) 457. 
%================================================= 
\bibitem{Diakonov:1995qy} 
D.~Diakonov, M.~V.~Polyakov and C.~Weiss, Nucl.\ Phys.\ B {\bf 461} 
(1996) 539.   
%================================================= 
\bibitem{Diakonov:2002fq} 
D.~Diakonov, Prog.\ Part.\ Nucl.\ Phys.\  {\bf 51} (2003) 173. 
%================================================= 
\bibitem{Musakhanov:1998wp} 
M.~Musakhanov, Eur.\ Phys.\ J.\ C {\bf 9} (1999) 235. 
%================================================= 
\bibitem{Musakhanov:2001pc} 
M.~Musakhanov, hep-ph/0104163. 
%================================================= 
\bibitem{Musakhanov:vu} 
M.~Musakhanov, Nucl.\ Phys.\ A {\bf 699} (2002) 340. 
%================================================= 
\bibitem{Nowak:1988bh} 
  M.~A.~Nowak, J.~J.~M.~Verbaarschot and I.~Zahed, 
  %``Flavor Mixing In The Instanton Vacuum,'' 
  Nucl.\ Phys.\ B {\bf 324} (1989) 1. 
%================================================= 
\bibitem{Alkofer:1989uj} 
  R.~Alkofer, M.~A.~Nowak, J.~J.~M.~Verbaarschot and I.~Zahed, 
  %``Pseudoscalars In The Instanton Liquid Model,'' 
  Phys.\ Lett.\ B {\bf 233} (1989) 205. 
%================================================= 
\bibitem{Kacir:1996qn} 
  M.~Kacir, M.~Prakash and I.~Zahed, 
  %``Hadrons and QCD instantons: A bosonized view,'' 
  Acta Phys.\ Polon.\ B {\bf 30} (1999) 287. 
%================================================= 
\bibitem{Pobylitsa:1989uq} 
P.~V.~Pobylitsa, Phys.\ Lett.\ B {\bf 226} (1989) 387. 
%================================================= 
\bibitem{Kim:2005jc} 
H.-Ch.~Kim, M.~M.~Musakhanov and M.~Siddikov, Phys.\ Lett.\ B {\bf 633} 
(2006) 201.   
 %================================================= 
\bibitem{Schafer:1996wv} 
T.~Schafer and E.~V.~Shuryak, Rev.\ Mod.\ Phys.\  {\bf 70} (1998) 323. 
%================================================= 
\bibitem{Jamin:2002ev} 
M.~Jamin, Phys.\ Lett.\ B {\bf 538} (2002) 71. 
%================================================= 
\bibitem{DiGiacomo:2004ff} 
A.~Di Giacomo and Y.~A.~Simonov, Phys.\ Lett.\ B {\bf 595} 
(2004) 368.  
%================================================= 
\bibitem{Beneke:1992ba} 
M.~Beneke and H.~G.~Dosch, Phys.\ Lett.\ B {\bf 284} (1992) 116. 
%================================================= 
\bibitem{Aladashvili:1995zj} 
K.~Aladashvili and M.~Margvelashvili, Phys.\ Lett.\ B {\bf 372} (1996) 299.  
%================================================= 
\bibitem{Khatsimovsky:1987bb} 
V.~M.~Khatsimovsky, I.~B.~Khriplovich and A.~R.~Zhitnitsky, Z.\ Phys.\ 
C {\bf 36} (1987) 455.  
%================================================= 
\bibitem{Ovchinnikov:1988gk} 
A.~A.~Ovchinnikov and A.~A.~Pivovarov, Sov.\ J.\ Nucl.\ Phys.\  {\bf 48} 
(1988) 721 [Yad.\ Fiz.\  {\bf 48} (1988) 1135]. 
%================================================= 
\bibitem{Braun:2004vf} 
  V.~M.~Braun and A.~Lenz, 
  %``On the SU(3) symmetry-breaking corrections to meson distribution 
  %amplitudes,'' 
  Phys.\ Rev.\ D {\bf 70} (2004) 074020. 
%================================================= 
%\cite{Nam:2006au} 
\bibitem{Nam:2006au} 
S.~i.~Nam, H.-Ch.~Kim, A.~Hosaka and M.~M.~Musakhanov, 
%  ``The leading-twist pion and kaon distribution amplitudes from the QCD 
 %instanton vacuum,'' 
Phys.\ Rev.\ D {\bf 74} (2006) 014019. 
  %%CITATION = HEP-PH 0605259;%% 
%================================================= 
\end{thebibliography}
\end{document}